\documentclass[aps,prb,showpacs,twocolumn,groupedaddress]{revtex4}

\usepackage[T1]{fontenc}
\usepackage[latin1]{inputenc}
\usepackage{latexsym}
\usepackage{array}
\usepackage{amsfonts}
\usepackage[centertags]{amsmath}
\usepackage{bm}
\usepackage{wasysym}
\usepackage[dvips,xdvi]{graphicx}
\usepackage{textcomp}
\usepackage{epsf}
\usepackage{hhline}

\newenvironment{eqnabc}{\begin{subequations}\begin{eqnarray}}{\end{eqnarray}\end{subequations}}

\begin{document}

\title{Confinement effects on glass forming liquids probed by DMA}

\author{J.~Koppensteiner}\email{johannes.koppensteiner@univie.ac.at}

\author{W.~Schranz}\email{wilfried.schranz@univie.ac.at}

\author{M.~R.~Puica}\email{madalina-roxana.puica@univie.ac.at}

\affiliation{Faculty of Physics, University of Vienna,
Boltzmanngasse\,5, A-1090 Vienna, Austria}

\date{\today}

\begin{abstract}
\noindent
Many molecular glass forming liquids show a shift of the glass transition $T_g$ to lower temperatures when the liquid is confined into mesoporous host matrices. Two contrary explanations for this effect are given in literature: First, confinement induced acceleration of the dynamics of the molecules leads to an effective downshift of $T_g$ increasing with decreasing pore size. Secondly, due to thermal mismatch between the liquid and the surrounding host matrix, negative pressure develops inside the pores with decreasing temperature, which also shifts $T_g$ to lower temperatures. Here we present novel dynamic mechanical analysis measurements of the glass forming liquid salol in Vycor and Gelsil with pore sizes of $d=$ 2.6, 5.0 and 7.5 nm. The dynamic complex elastic susceptibility data can be consistently described with the assumption of two relaxation processes inside the pores: A surface induced slowed down relaxation due to interaction with rough pore interfaces and a second relaxation within the core of the pores. This core relaxation time is reduced with decreasing pore size $d$, leading to a downshift of $T_g\propto1/d$ in perfect agreement with recent DSC measurements. Thermal expansion measurements of empty and salol filled mesoporous samples revealed that the contribution of negative pressure to the downshift of $T_g$ is small (<\;30\%) and the main effect is due to the suppression of dynamically correlated regions of size $\xi$ when the pore size $d$ approaches $\xi$.

\end{abstract}

\maketitle

\section{Introduction}

When approaching a glass transition some physical properties like viscosity or relaxation times change up to 14 orders of magnitude\cite{cukiermann,ediger}. An explanation for the observed slowing down of the dynamics
is the formation of collectively dynamically rearranging clusters \cite{begriff_rearranging, begriff_collectively_rearranging} or regions, with growing size $\xi$ and increasing relaxation times as $T_g$ is approached\cite{jaeckle}. The idea of an increasing dynamic correlation length $\xi$ when approaching a glass transition is strongly supported by recent computer simulations \cite{Donati,Bennemann,scheidler3}, although not strictly proven, since computer simulations cannot treat the time range of the $\alpha$-process.
Very recently a breakthrough was achieved in this field. G.~Biroli, et al.~\cite{Biroli} found first-time evidence that the Mode Coupling Theory predicts a growing dynamic length scale approaching the glass transition of a supercooled liquid. The authors obtained a rather modest growth of the dynamical length scale $\xi$ with decreasing temperature, which is in very good agreement with computer simulations \cite{scheidler3} and experimental results. Indeed many experimental setups like heat capacity spectroscopy \cite{hempel_xiauscp,Donthhuth}, multidimensional NMR \cite{Tracht,Reinsberg,Qiu}, multipoint dynamical susceptibilities \cite{Berthier}, etc. have been used to monitor a possible growing length scale accompanying the glass transition. All these results agree in the fact, that the obtained dynamically correlated regions - although material dependent - are of the order of 1-4 nm and display - if at all - a weak temperature dependence.

\qquad An alternative experimental approach to get a reference to a possibly existing cooperation length $\xi$ which increases when $T \rightarrow T_g$ is by spatial limitation of a glass forming liquid. Spatially confining geometries as ultrathin films, mesoporous silica or zeolithes have already been used to study phase transitions of water\cite{water}, hydrocarbons\cite{hydrocarbons}, noble gases\cite{noblegases, noblegases2}, liquid crystals\cite{liquidcrystals} or alkenes \cite{alkenes}. But this concept also illuminated the old and still open question on the very nature  of the glass transition and its dynamics\cite{glasstransition}: In a pioneering work Jackson and McKenna\cite{calorimetric2} studied the glass transition of organic liquids in controlled pore glasses (CPG) for various pore sizes $d$. They found a reduction of the glass transition temperature $T_g$ for liquids in confinement as compared to the bulk material. The downshift of $T_g$ was larger for smaller pore sizes, i.e.~$\Delta T_g \propto 1/d$, an effect similar - but not as strong as - the suppression known for the melting temperature $T_m$ in confinement. During the following two decades this effect was studied via calorimetry\cite{trofymluk,calorimetric3}, dielectric spectroscopy \cite{dielectric_overview}, neutron scattering\cite{neutron}, light scattering\cite{patkowski} and molecular dynamics\cite{scheidler1}. It was shown that in many cases confinement below a characteristic length impedes\cite{calorimetric3} the transition, implying that molecules within a region of the size $\xi_g$ (approaching $T_g$ typically some nm\cite{sillescu,donthbuch}) have to cooperate and rearrange in order to create the glassy state. Hindering this cooperation first leads to a downshift of $T_g$ if $d\sim\xi_g$ and finally to a suppression of the transition if $d<\xi_g$\cite{calorimetric3}.

\qquad However, although this shift of $T_g$ with decreasing confinement size was found in many studies, there are complications which blur this simple picture: E.g.~in some systems a competition appears between slowing down of molecular motions due to pinning of the molecules at the pore surface and acceleration of the dynamics due to decreasing size of the confinement. Another effect occurs due to the difference in thermal expansion coefficients of the porous host matrix and the glass forming liquid. This may create negative pressure upon the confined liquid when the glass transition is approached. Some authors attribute the whole observed downshift of $T_g$ to this negative pressure effect \cite{patkowski}. We will address these points in very detail below. For excellent reviews about these topics the reader is referred to Refs.~\onlinecite{review_Alcoutlabi} and \onlinecite{Simionesco}.

\qquad Very recently the confinement effect on the glass-forming liquid salol was studied via dynamic mechanic analysis measurements (DMA)\cite{Schranz1} in Vycor with $d=7$ nm pore size. It turned out that the dynamic elastic response is very sensitive to the glass transition of liquids confined to mesoporous samples. Based on the results of computer simulations\cite{scheidler1, scheidler2} we could disentangle acceleration effects due to confinement and  slowing down of molecular motion due to interaction of the molecules with the rough pore surface.
We could even predict the pore size dependence of the dynamic elastic response (see Fig.~4 of Ref.~\onlinecite{Schranz1}). In order to test these predictions and to study the glass transition of salol for different pore sizes, further measurements have been performed. Here we present novel experimental results of the temperature and frequency dependence of the complex dynamic elastic susceptibility of salol confined in mesoporous matrices of $d=7.5$ nm, 5.0 nm and 2.6 nm. In addition, thermal expansion measurements have been performed, which now allows us to take a new look at the often discussed negative pressure effect on glass forming liquids in confinement and to seperate this effect from an intrinsic size effect.

\qquad The present paper is organized as follows: Section II yields insight into sample preparation and some technical details of DMA analysis. Section III displays a compilation of the experimental data and results of modeling and interpretation of the present data. It also contains a calculation of the effect of adsorption swelling and the separation of the actual downshift of $T_g$ in salol into the negative pressure effect and the confinement effect. Section IV concludes the paper.

\section{Experimental}

\subsection{Sample preparation}

Porous Vycor samples are made by Corning Inc., NY, and sold under the brand name "Vycor 7930". Via phase separation and leaching a three dimensional random network of pores in nearly pure silica is fabricated\cite{elmer}. Pores are uniformly distributed in length, direction and density\cite{Levitz}. The mean ratio of average pore diameter $d$ and pore length $l$, is $d/l\approx0.23$. Gelsil is a mesoporous xerogel consisting of pure silica with a very narrow pore radius distribution. Gelsil rods were made by 4F International Co., Gainesville, FL. Results on pore sizes derive from BJH analysis of the individual $\textrm{N}_2$-desorption isotherms\cite{Rouguerol} and are summarized in Tab.~\ref{tab:samples_data}.

\begin{table}[h!]
\caption{$\textrm{N}_2$ adsorption characteristics of porous silica samples.} \label{tab:samples_data}
\begin{center}
\begin{tabular}{llll}
\qquad&\qquad Gelsil2.6&\qquad Gelsil5&\qquad Vycor\\
\hline \hline
av. pore diameter\;({nm})&\qquad 2.6 &\qquad 5.0 &\qquad 7.5 \\
surface area\;($\textrm{m}^2$/g)&\qquad 586 &\qquad$509$&\qquad 72\\
pore volume\;($\textrm{cm}^3$/g)&\qquad 0.376 &\qquad 0.678 &\qquad 0.214 \\
porosity $\phi$\;&\qquad 0.51 &\qquad 0.66 &\qquad 0.30\\
\hline \hline
\end{tabular}
\end{center}
\end{table}

\qquad All samples were cut and sanded in order to gain parallel surface plains. The typical size of a sample was $(2\times2\times8)\;\textrm{mm}^3$ for parallel plate and about $(2\times1\times7)\;\textrm{mm}^3$ for three point bending DMA-measurements. Cleaning was done in a 30\% hydrogen peroxide solution at $90^\circ$C for 24 h, drying at $120^\circ$ C in a high-vacuum chamber at $10^{-6}$ bar, also for about 24 h. The guest glass forming material was salol (phenyl salicylate, $\textrm{C}_{13}\textrm{H}_{10}\textrm{O}_3$), a low molecular weight liquid, whose melting temperature is $T_m=316$~K. Salol is a standard, so called fragile\cite{fragility}, glass former (m=73) known\cite{trofymluk} to form a glass either at extreme cooling rates of 500 K/min or in pores smaller than 11.8~nm. Filling was done at 317 K via capillarity wetting. By comparing the weight of clean and filled samples the filling fractions $f$ were determined (see Table III).

\subsection{Dynamic mechanical analysis (DMA)}
\begin{figure}
\begin{center}
\includegraphics[scale=0.5]{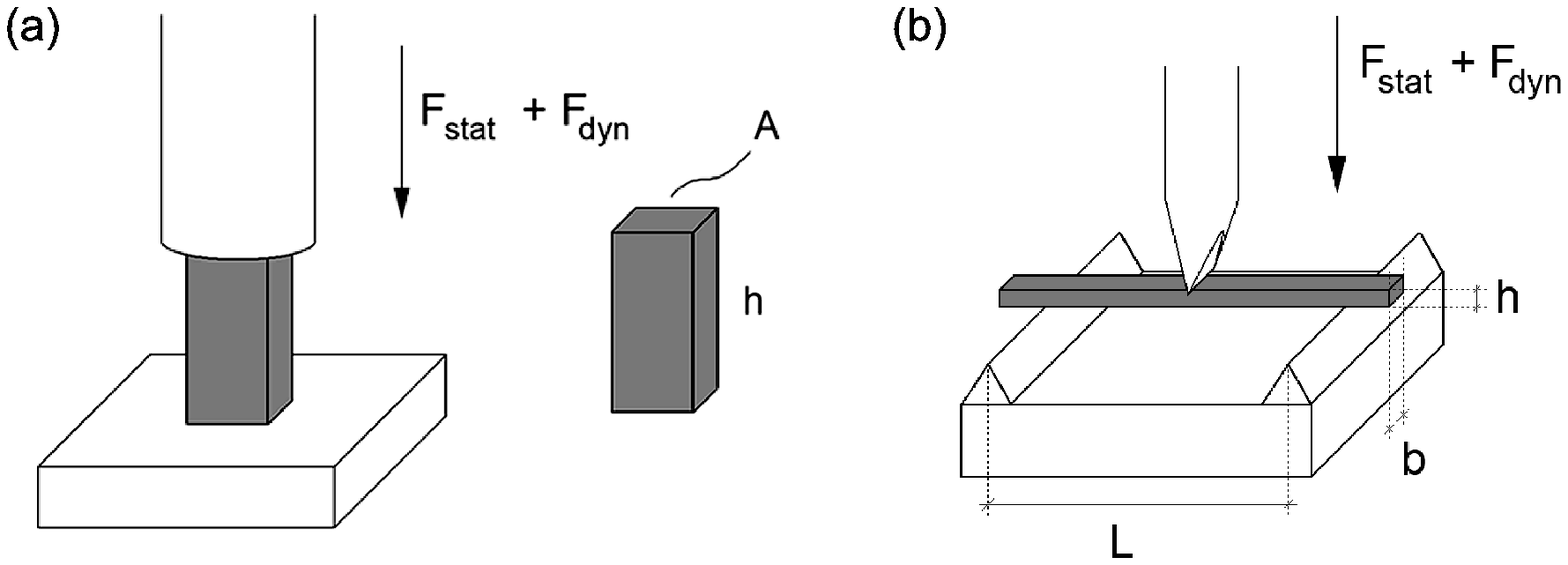}
\caption{Sketch of (a) parallel plate and (b) three point bending geometry.}\label{fig:3pb_pp_skizze}
\end{center}
\end{figure}

In this method a static and a dynamic force $F_{stat}+F_{dyn}\cdot e^{i\omega t}$ (0.001-16 N at 0.01-100 Hz) are applied on a sample using a quartz or steel rod (see Fig.~\ref{fig:3pb_pp_skizze}). The response of the sample is measured via the displacement of the rod. Absolute height $h$, height amplitude $\Delta h$ and phase lag $\delta$ are read via electromagnetic inductive coupling (LVDT) with a resolution of 10~nm and $0.01^\circ$ respectively. These data allow direct access to real and imaginary part of the complex elastic susceptibility at low frequency and as a function of temperature and applied force. In addition, the thermal expansion of a sample can be determined in the so called TMA-mode, where no external force is applied. Two devices are used: a DMA 7  and a Diamond DMA, both from Perkin Elmer Inc. Two measuring geometries are applied: parallel plate compression (PP) and three-point bending (3PB), see Fig.~\ref{fig:3pb_pp_skizze}.

\qquad Parallel plate geometry reveals purely the complex Young´s modulus $Y^{*}=Y'+iY''$, where $Y'$ and $Y''$ are the storage and the loss modulus, respectively. The three point bending geometry delivers Young´s modulus plus a small (geometry dependent) contribution of a shear elastic constant. More details on measurement geometry may be found in Refs.~\onlinecite{dmaold1} and \onlinecite{dmaold2}. The absolute accuracy of resulting real and imaginary parts $Y'$ and $Y''$ is rather poor, mainly because of contact losses between the quartz rod and the sample. A discussion of these systematic errors may be found in Ref.~\onlinecite{Schranz1}. In contrast the relative accuracy is excellent and the DMA-method is estimated to be about 100 times more sensitive to detect glass transitions or other subtle phase transitions than DSC-measurements \cite{Menard}.

\section{Results and Discussion}

\subsection{Dynamic elastic response}

\begin{figure}
\begin{center}
\includegraphics{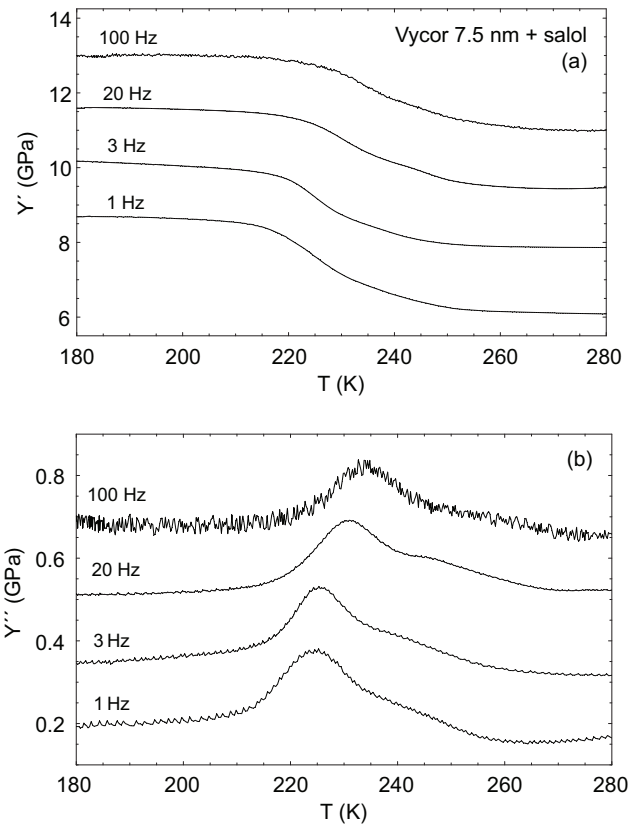}
\caption{Real (a) and imaginary parts (b) of the complex Young´s modulus of Vycor 7.5 nm filled with salol (filling fraction $f \approx 0.79$) measured in three point bending geometry. The curves are offset from the 1 Hz data for sake of clarity.}\label{fig:Vycorfilled_3pb_diamond}
\end{center}
\end{figure}

\begin{figure}
\begin{center}
\includegraphics{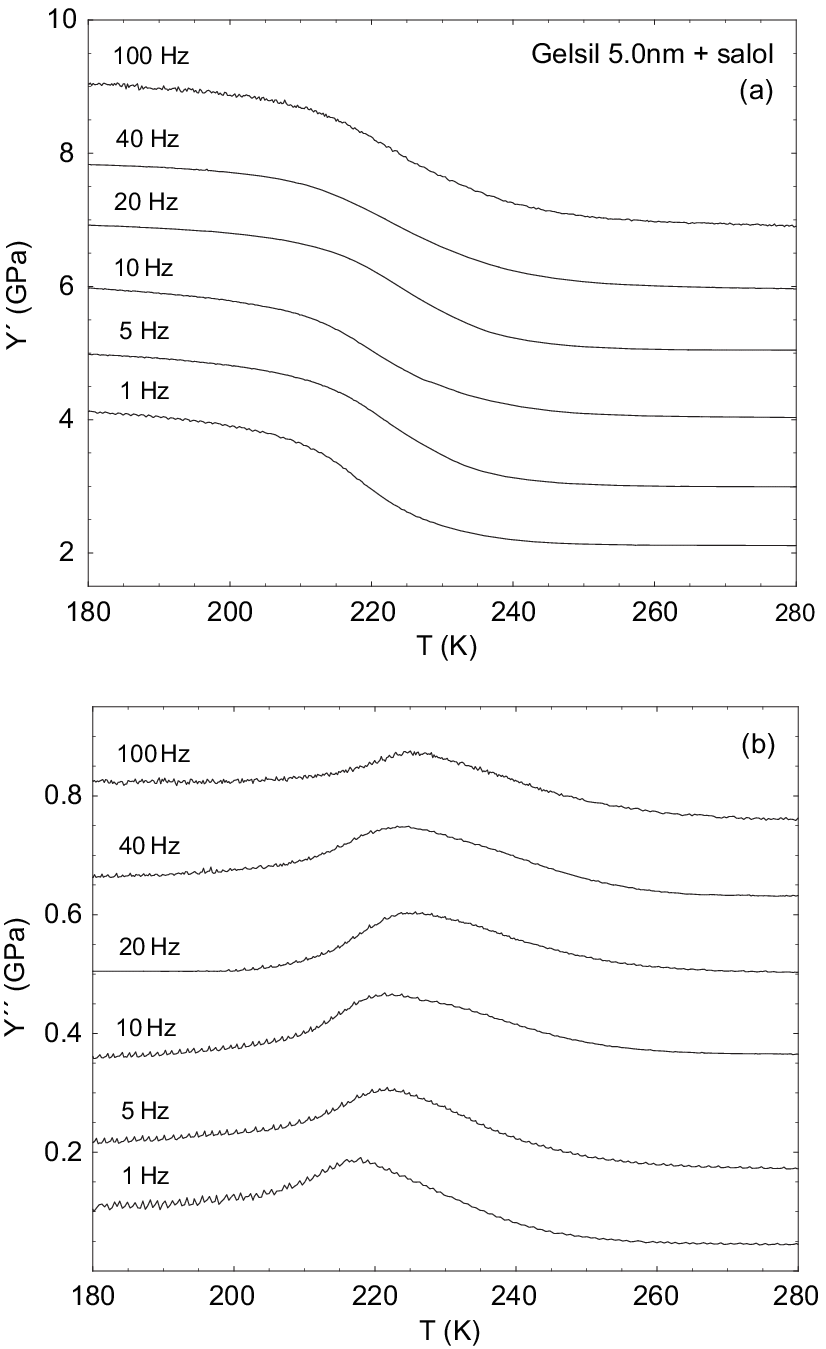}
\caption{Real (a) and imaginary (b) parts of the complex Young´s modulus of
Gelsil 5.0 nm filled with salol (filling fraction $f \approx 0.75$)
measured in parallel plate geometry (Diamond DMA). 1Hz signal are original data, other signals are offset for sake of clarity.}\label{fig:Gelsil5nm_filled_pp_overview}
\end{center}
\end{figure}

\begin{figure}
\begin{center}
\includegraphics{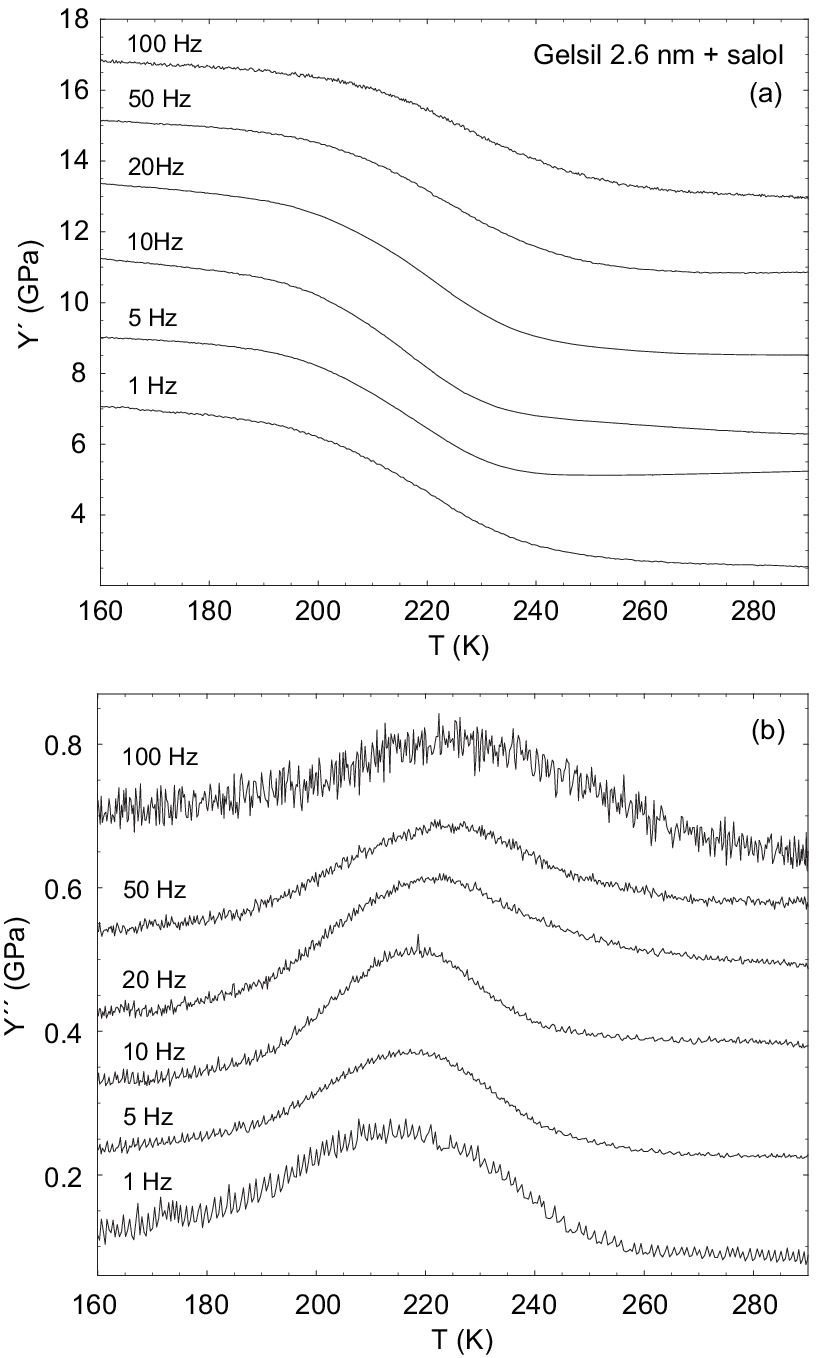}
\caption{Real (a) and imaginary (b) parts of the complex Young´s modulus of
Gelsil 2.6 nm filled with salol (filling fraction $f \approx 0.65$)
measured in three point bending geometry (Diamond DMA). 1 Hz signal are original data, other signals are offset for sake of clarity.}\label{fig:Gelsil2.6nm_filled_3pb_overview}
\end{center}
\end{figure}

\begin{figure*}
\begin{center}
\includegraphics{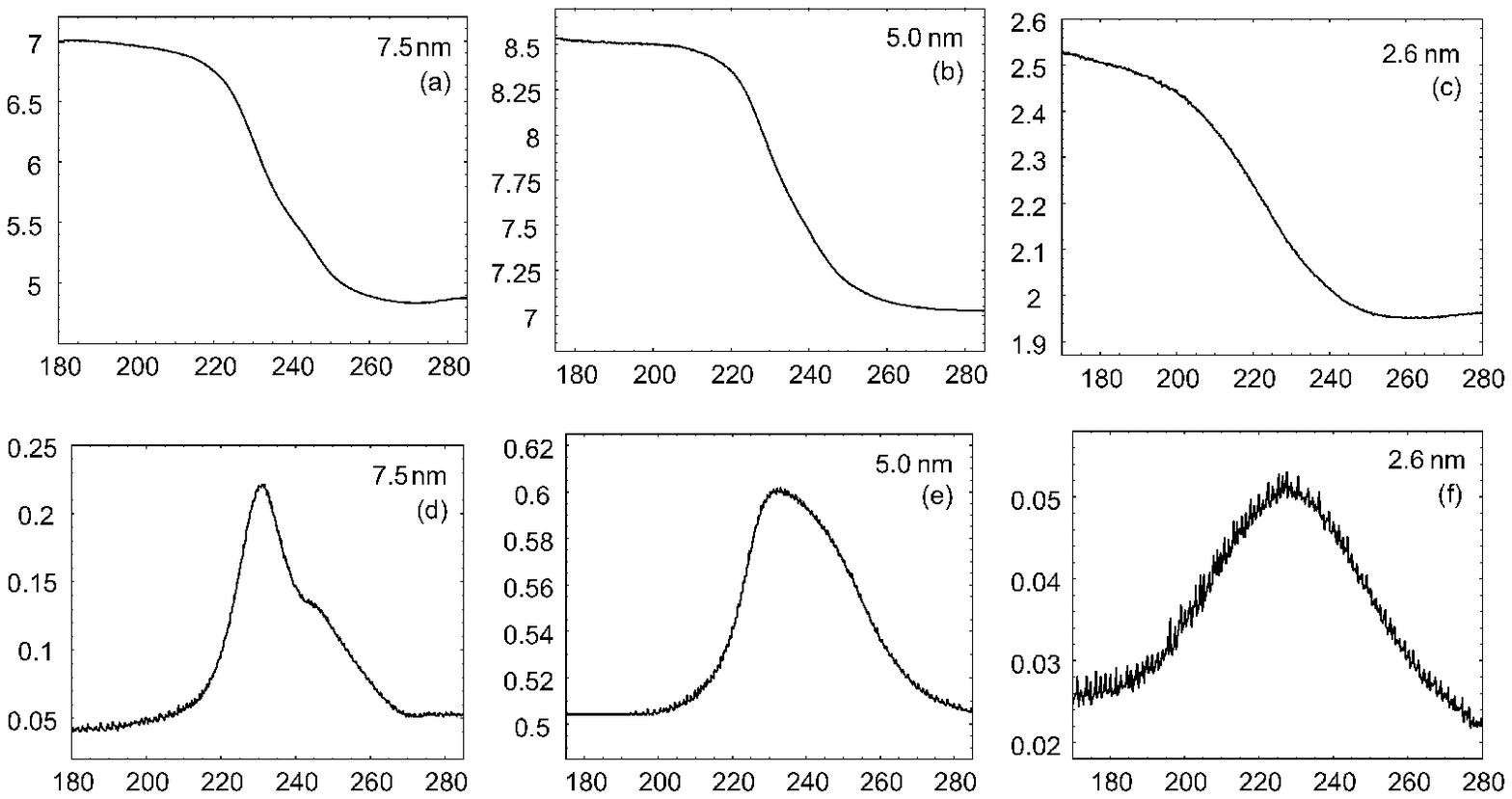}
\caption{Real (a) and imaginary (b) part of the complex Young´s modulus of salol in Vycor or Gelsil for different pore sizes, all measured at 20 Hz.} \label{fig:real_imag_pores_overview}
\end{center}
\end{figure*}

Diamond DMA-measurements (in parallel plate and three-point-bending geometry) of Vycor and Gelsil samples filled with salol  are shown in Figs.~\ref{fig:Vycorfilled_3pb_diamond} - \ref{fig:real_imag_pores_overview}. The loss modulus $Y''$ (Fig.~\ref{fig:Vycorfilled_3pb_diamond}b) of salol in 7.5 nm pores clearly shows a "two peak structure", i.e.~a peak with HWHM about 20 K, and a shoulder or second peak at about 15 K higher temperature (also see Fig.~\ref{fig:real_imag_pores_overview}b). This is also reflected by the real part $Y'$, which displays a "two step like shape" with temperature (Figs.~\ref{fig:Vycorfilled_3pb_diamond}a and \ref{fig:real_imag_pores_overview}a).
Both peaks in $Y''$ shift to higher temperatures with increasing frequency as expected for a glass transition. In smaller pores of Gelsil 5.0, peak and shoulder merge into one asymmetric peak of width $\sim$  30 K (see Fig.~\ref{fig:Gelsil5nm_filled_pp_overview}b and \ref{fig:real_imag_pores_overview}e), also shifted with higher frequency to higher temperatures. In 2.6 nm pores the loss peak shows a rather symmetric form broadened up to about 60 K, see Figs.~\ref{fig:Gelsil2.6nm_filled_3pb_overview}b and \ref{fig:real_imag_pores_overview}f.

\qquad While in large pores of 7.5 nm diameter vitrification of salol seems to happen decoupled (two peaks in $Y''$) in regions near the pore surface and the pore center, things change in smaller pores. With decreasing pore diameter, $Y''$ approaches a symmetric form and simultaneously $Y'$ changes from a "double step shape" into a "single step shape", indicating only one type of relaxation process. Similar broadening effects as for the loss peaks of our DMA-measurements were observed in pores of decreasing size also by calorimetric\cite{calorimetric2} and dielectric measurements\cite{relax_peak_broadening, Pissis}. This broadening as well as a shift of the glass transition to lower temperatures was calculated by Sappelt and Jäckle using kinetic Ising and lattice gas models\cite{Sappelt}, and shown to originate from confinement induced suppression of cooperative motion of molecules.

\qquad Pure Vycor and Gelsil, meaning exposed to air and therefore mostly filled with nitrogen, do not show any of these features. $Y'$ decreases about 2\% between 300 K and 180 K. $Y''$ is constant within the corresponding temperature range.\newline

Any standard relaxation model like Debye, Kohlrausch, Cole-Cole or Cole-Davidson fails to describe our dynamic elastic susceptibility data
if only one type of relaxation process is assumed. One would have to use extreme stretching parameters to fit $Y'$, which then leads to improper temperature shifts of the peaks in $Y''$ with respect to the experimental data and misfitting signal heights. The most efficient model to describe our data turned out to be a modification of the empirical Vogel-Fulcher-Tammann-law

\begin{equation}
\tau(T)=\tau_0\cdot exp[\frac{E}{T-T_0}]\;,
\label{eqn:VFT_temp}
\end{equation}

where $\tau_0$ is a preexponential factor, $E\cdot k_B$ is an activation energy, and $T_0$ is the Vogel-Fulcher (VF) temperature. Following computer simulations\cite{scheidler1, scheidler2} we take into account a shift of VF-temperatures along the pore radius $r$ . In a recent paper Zorn et al.~\cite{zorn1} suggest the empirical ansatz

\begin{equation}
T_0(r)=T_{00}+\frac{k}{R-r+r_p}\;,
\label{eqn:VFTshift_zorn}
\end{equation}

with the bulk VF-temperature $T_{00}$, and the pore radius $R=d/2$. The so called penetration radius $r_p$ is the radius beyond which it is very unlikely to find a particle in the fluid state\cite{scheidler1}. The combination of Eqns.~(\ref{eqn:VFT_temp}) and (\ref{eqn:VFTshift_zorn}) leads to a radial distribution of relaxation times $\tau$ inside the pore:

\begin{equation}
\tau(r,T)=\tau_0 \cdot exp\big[\frac{E}{T-(T_{00}+\frac{k}{R-r+r_p})}\big]
\label{eqn:tau_r_T}
\end{equation}

Eqn.~(\ref{eqn:tau_r_T}) describes the exponential increase of relaxation time when a rough pore wall is approached, and a growing influence of the pore wall with decreasing temperature, a behaviour which was also found by recent computer simulations\cite{scheidler1,scheidler2,scheidler3,scheidler4}. A Cole-Cole-Plot of $log(Y')$ vs.~$log(Y'')$ of our data clearly calls for a Cole-Davidson model of the complex dynamic elastic susceptibility

\begin{equation}
Y^{*}(\omega)\propto \frac{1}{(1+i\omega \tau)^{\frac{\gamma}{2}}}
\label{eqn:cole-davidson}
\end{equation}

with $\omega=2 \pi \nu$, $\nu$ being the measurement frequency, and the broadening parameter $\gamma$. Using Eqn.~(\ref{eqn:tau_r_T}), averaging over the pore radius R, and separating real and imaginary part of $Y^{*}=Y'+i\cdot Y''$ by common procedures leads to

\begin{figure*}
\begin{center}
\includegraphics{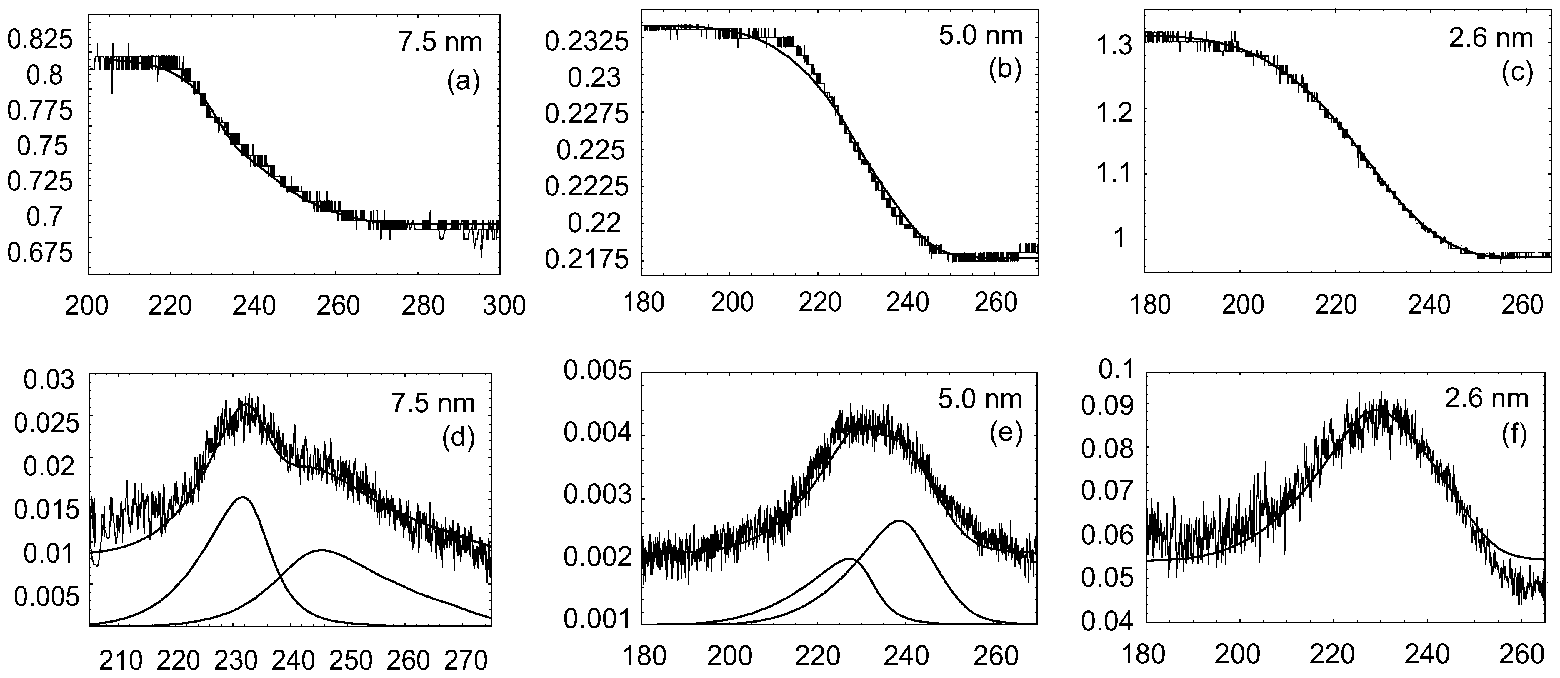}
\caption{Real part $Y'$ and imaginary part $Y''$ of different porous samples filled with salol. Lines are fits using Eqns.~(5) with parameters of Tab.~\ref{tab:FitparameterVycorGelsil}.} \label{fig:pores_fit_overview}
\end{center}
\end{figure*}

\begin{eqnabc}
Y'=1-\frac{2\;\Delta Y}{R^2}\int_0^R \frac{cos[\gamma\cdot arctan(\omega\tau(r,T))]}{[1+\omega^2\tau(r,T)^2]^\frac{\gamma}{2}}r dr\\
Y''=\frac{2\;\Delta Y}{R^2}\int_0^R \frac{sin[\gamma\cdot arctan(\omega\tau(r,T))]}{[1+\omega^2\tau(r,T)^2]^\frac{\gamma}{2}}r dr
\label{eqn:YstrichYzweistrichlang}
\end{eqnabc}
As already mentioned above, the two peak structure in $Y''$ of 7.5~nm and 5~nm confined salol (Figs.~\ref{fig:Vycorfilled_3pb_diamond} and \ref{fig:Gelsil5nm_filled_pp_overview}) suggests to split the dynamic elastic response into a core and a surface contribution: the molecules in the core (center of the pores) behave bulk-like and are dynamically decoupled from the molecules near the pore surface.
This is modelled by inserting into Eqn.~(\ref{eqn:YstrichYzweistrichlang}) the corresponding relaxation times $\tau_0\cdot exp(E/(T-T_0))$ given by Eqn.~(\ref{eqn:VFT_temp}) if $r \leq R_c$ and $\tau(r,T)$ given by Eqn.~(\ref{eqn:tau_r_T}) if $r>R_c$ (see also Fig.~7).
The sum of the two contributions perfectly describes our $Y'$ and $Y''$ data on salol in 7.5~nm and 5~nm pores simultaneously (see Fig.~\ref{fig:pores_fit_overview}).

\qquad In 2.6 nm pores we do not expect any molecule to behave like the bulk liquid any more, since the pore radius is of the same order as the estimated surface shell (see Tab.~\ref{tab:FitparameterVycorGelsil}), implying that every molecule is influenced by the near surface. Therefore we use Eqns.~(5) with no bulk term which reproduces one single peak and also fits our data very well (Fig.~\ref{fig:real_imag_pores_overview}c and f and Fig.~\ref{fig:pores_fit_overview}c and f).

\quad The radius of the "core" of bulk-like interacting molecules turned out to be $R_c=$~2.5 nm and 1.35~nm in 7.5~nm and 5.0~nm pores, respectively, see Tab.~\ref{tab:FitparameterVycorGelsil}. This implies that the thickness of the shell of molecules being slowed down by wall interaction $R-R_c= 1.25$~nm and 1.15~nm for 7.5~nm and 5.0~nm pores respectively.

\begin{figure*}
\begin{center}
\includegraphics{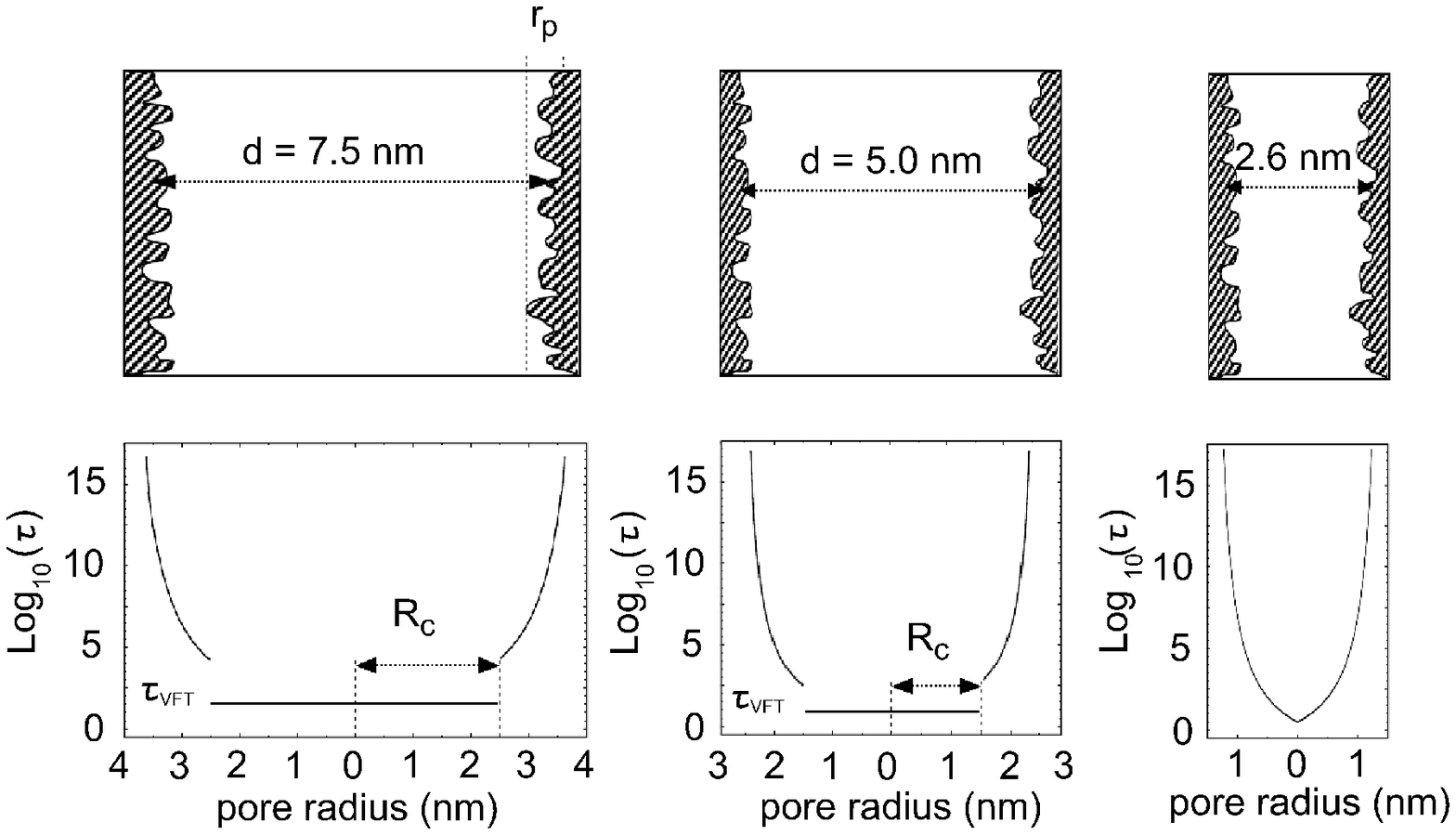}
\caption{Modelled relaxation time distributions in pores of diameter 7.5~nm to 2.6~nm from Eqn.~(\ref{eqn:tau_r_T}) used in Eqns.~(5) for fits of Fig.~\ref{fig:pores_fit_overview}.} \label{fig:pore_model}
\end{center}
\end{figure*}

\qquad Additional loss peaks, attributed to molecules forming H-bonds to the inner pore surface, have also been reported from dielectric measurements of salol in 7.5~nm pores\cite{dielectric1,dielectric2}. The work of Kremer and Stannarius\cite{dielectric1} also revealed that the typical size of a shell of molecules interacting with the pore surface is about 2 or 3 molecules. Since the size of a salol molecule is estimated as $(1.4\times0.6\times0.4)\;\textrm{nm}^3$ in Ref.~\onlinecite{salolsize} or as $0.282\;\textrm{nm}^3$ in Ref.~\onlinecite{salolsize2}, both corresponding to a mean diameter of 0.8~nm, this shell size is in the order of 1.6 to 2.4~nm. This is in very good agreement with our findings (see Tab.~\ref{tab:FitparameterVycorGelsil}). The core size $R_c$ decreases with decreasing pore radius (see Tab.~\ref{tab:FitparameterVycorGelsil} and Fig.~\ref{fig:pore_model}), also in very good agreement with the results of Kremer et al.

\begin{table}[h!]
\caption{Fit parameters used in Eqns.~(5) for fits of Fig.~\ref{fig:pores_fit_overview}.} \label{tab:FitparameterVycorGelsil}
\begin{center}
\begin{tabular}{llll}
\qquad&\qquad Vycor&\qquad Gelsil5&\qquad Gelsil2.6\\
\hline \hline
$R$\;(nm)&\qquad$3.75$&\qquad$2.50$&\qquad$1.28$\\
$r_p$\;(nm)&\qquad$0.36$&\qquad$0.25$&\qquad$0.28$\\\hline
$E$\;(K)&\qquad$1750$&\qquad$1750$&\qquad$1750$\\
$T_{00}$\;(K)&\qquad$158.5$&\qquad$156.2$&\qquad$136.0$\\
$\tau_0$\;(s)&\qquad$10^{-11}$&\qquad$10^{-11}$&\qquad$10^{-11}$\\
$\gamma$&\qquad$0.33$&\qquad$0.18$&\qquad$0.15$\\
$k$\;(nm$\cdot$ K)&\qquad$18$&\qquad$11$&\qquad$25$\\
$R_c$\;(nm)&\qquad 2.5&\qquad 1.35&\qquad -\\
shell $R-R_c$(nm)\;&\qquad 1.25&\qquad 1.15&\qquad 1.28\\
\hline \hline
\end{tabular}
\end{center}
\end{table}

\begin{figure}
\begin{center}
\includegraphics[scale=0.8]{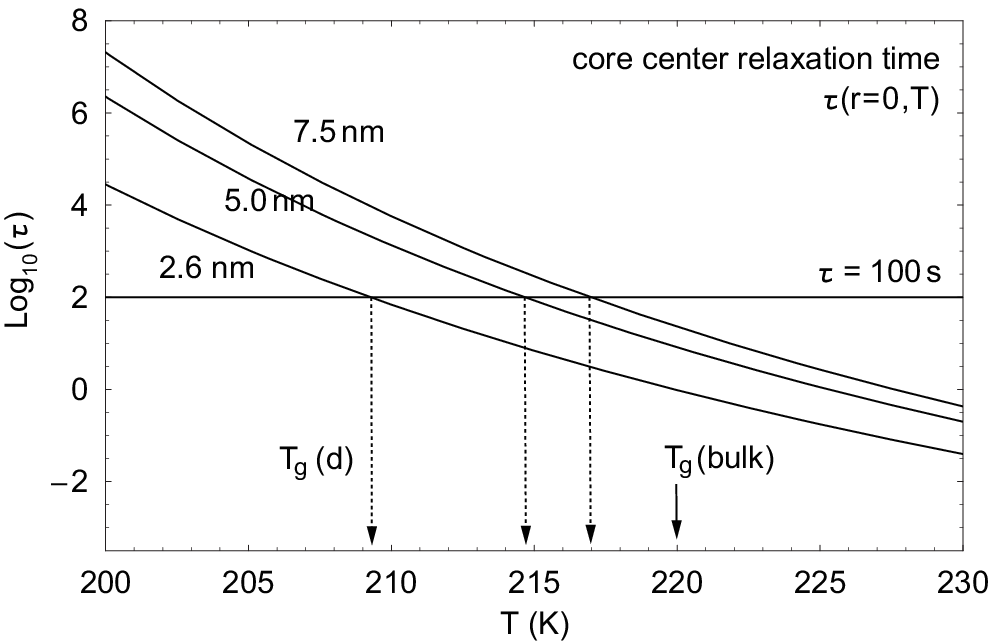}
\caption{Relaxation time in pore centers calculated from Eqns.~(\ref{eqn:VFT_temp}) and (\ref{eqn:tau_r_T}) with corresponding parameters from Tab.~\ref{tab:FitparameterVycorGelsil}. Horizontal line shows $\tau=100$ s.} \label{fig:pore_center_tau}
\end{center}
\end{figure}

The fitted Vogel-Fulcher temperature $T_{00}$ is reduced with respect to the bulk and with decreasing pore size (see Tab.~\ref{tab:FitparameterVycorGelsil}). In order to compare our results with published data, we plotted the relaxation time in the pore center $\tau(r=0,T)$ for various pore sizes $d$ and determined the corresponding $T_g(d)$ by using the common procedure\cite{Tg_def_Richert} for finding the laboratory glass transition temperature, i.e.~a cut with a horizontal line at $\tau=100$~s (see Fig.~\ref{fig:pore_center_tau}). As shown in Fig.~\ref{fig:DeltaTg_pore diameter}, this leads to glass transition temperatures decreasing $\propto 1/d$ in very good agreement with published data of DSC measurements\cite{patkowski}.

\begin{figure}
\begin{center}
\includegraphics[scale=1.08]{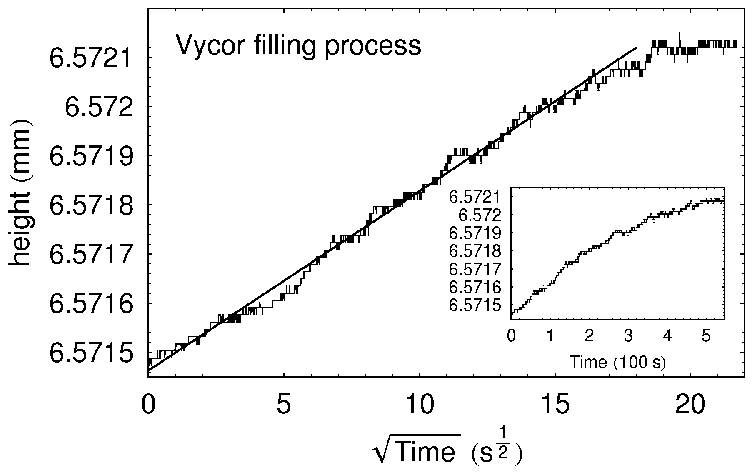}
\caption{Height of Vycor sample during the filling process against $\sqrt{t}$. Inset shows sample height against time.}
\label{fig:vycor_fillingprocess}
\end{center}
\end{figure}

\subsection{Filling process and accompanying effects}

By using a DMA in a static TMA mode one can detect small changes in a sample´s height with a resolution of 10 nm. We measured the time dependent swelling of the Vycor and Gelsil samples during filling with salol and the thermal expansion of empty and filled samples in the following way: In parallel plate mode, the quartz rod is placed on top of the sample with force $F=0$ N, and just height and temperature signals are read out. A clean piece of Vycor/Gelsil sample is cooled down to 170 K. Afterwards the sample is heated slightly above the melting temperature $T_m=316$ K of salol and kept there isothermally. Crystalline, powder-like salol placed right around the sample melts and percolates the Vycor/Gelsil sample due to capillarity (Fig.~\ref{fig:vycor_fillingprocess}). After filling until saturation, the sample is cooled down to 170 K again (Fig.~\ref{fig:filled_vs_clean_overview}).
\qquad The time-dependence of the filling process is displayed in Fig.~\ref{fig:vycor_fillingprocess} for Vycor. While salol is percolating the sample, the temperature is held constant and the sample´s height is measured. Charts for Gelsil 5.0 nm and Gelsil 2.6 nm look very similar. The diagrams in all cases show the typical  $\sqrt{t}$-behavior as expected for a single capillary rise experiment, following Lucas\cite{lucas} and Washburn\cite{washburn}. This result is in concordance with findings of P.~Huber et al.~\cite{rheology}, who investigated the mass uptake of porous silica samples and its time dependence, leading to the Lucas-Washburn $\sqrt{t}$ behaviour of the mass uptake with time. Very recently it was shown that the Lucas-Washburn equation (with small modifications) works well even at the nanoscale\cite{Dimitrov}, which is in harmony with our results.

\qquad The expansion of a porous sample during adsorption of gases or water has already been investigated in the 1920s\cite{Meethan}. As a liquid/gas intrudes the sample it is subject to a negative hydrostatic pressure inside the pores, which leads to an expansion of the porous sample during adsorption of gases or water. Mesoporous media have enormous inner surfaces up to some 100 $\textrm{m}^2$/g (see Tab.~\ref{tab:samples_data}). This leads to a considerable stress reduction within the whole matrix and a sudden voluminal growth, which slows down and stops as all pore space is filled (see Fig.~\ref{fig:vycor_fillingprocess}). The change in height due to the adsorption swelling can even be calculated quantitatively. The pressure reduction of the liquid in a capillary is known\cite{landau_lifshitz} as $P_c=2\sigma / r$, with the surface tension $\sigma$ and the capillary radius $r$. With $\sigma=1.73\cdot10^{-2}$ N/m from Ref.~\onlinecite{patkowski}, this yields a capillary pressure of 26.6 MPa in 2.6 nm pores. This would lead to a hypothetical capillary rise of 1.8 km for salol.
The linear strain $\epsilon=\Delta h /h$ accompanying the filling process can be computed by the equation\cite{drying_shrinkage}:

\begin{equation}
\epsilon=\frac{f\cdot P_c}{3}\Big(\frac{1}{K}-\frac{1}{K_s}\Big)
\label{eqn:filling_strain}
\end{equation}

with the filling fraction f, the bulk modulus $K$ of the empty host matrix, and the bulk modulus of the material building the solid frame $K_s$ (which is nearly pure $\textrm{SiO}_2$). The bulk moduli $K$ have been determined by resonant ultrasound spectroscopy (RUS)\cite{RUS_unpublished}. Tab.~\ref{tab:filling_exp} shows parameters used to calculate $\epsilon=\Delta h / h$. The calculated values for the adsorption swelling agree rather well with the experimental results.

\begin{table}[h!]
\caption{Variables of Eqn.~(\ref{eqn:filling_strain}).}  \label{tab:filling_exp}
\begin{center}
\begin{tabular}{llll}
\qquad&\qquad Vycor&\qquad Gelsil5&\qquad Gelsil2.6\\
\hline \hline
$d$ (nm)&\qquad 7.5&\qquad 5.0&\qquad 2.6 \\
porosity $\Phi$ &\qquad 0.31&\qquad 0.66 &\qquad 0.51\\
$P_c$ (MPa)&\qquad9.2&\qquad 13.8&\qquad26.6 \\
$K$ (GPa)&\qquad 8.1 &\qquad 3.9 &\qquad 9.6\\
$f$&\qquad 0.77 &\qquad 0.62 &\qquad 0.32\\ \hline
$\epsilon_{calc}$&\qquad$2.3\cdot10^{-4}$&\qquad $6.6\cdot10^{-4}$&\qquad$2.2\cdot10^{-4}$\\
$\epsilon_{exp}$&\qquad$1.0\cdot10^{-4}$&\qquad $4.1\cdot10^{-4}$&\qquad$3.5\cdot10^{-4}$\\
\hline\hline
\end{tabular}
\end{center}
\end{table}

\subsection{Negative pressure effect}

The downshift of the glass transition in nm-confining pores is often reported to obey a $1/d$ law, see Refs.~\onlinecite{calorimetric2}, \onlinecite{patkowski}, \onlinecite{trofymluk} and \onlinecite{neg_pressure1}. At first this was proposed by Jackson and McKenna\cite{calorimetric2}, following their former results on the shift of the melting transition $T_m$ in confinement\cite{Jackson_McKenna_Tmshift}. But the supposed suppression of molecular cooperation when the pore diameter approaches an inherent length scale is not the only possible reason for a downwards shift of $T_g$ in confinement. Zhang et al.\cite{neg_pressure1} proposed the increase of negative hydrostatic pressure within the pores due to mismatching thermal expansions of liquid and host matrix as the main driving force for the downshift of $T_g$. This idea is also discussed by Patkowski et al.\cite{patkowski} and Simon et al.\cite{enthalpy_recovery}, and was reviewed by Alcoutlabi and McKenna\cite{review_Alcoutlabi}.

\begin{figure*}
\begin{center}
\includegraphics{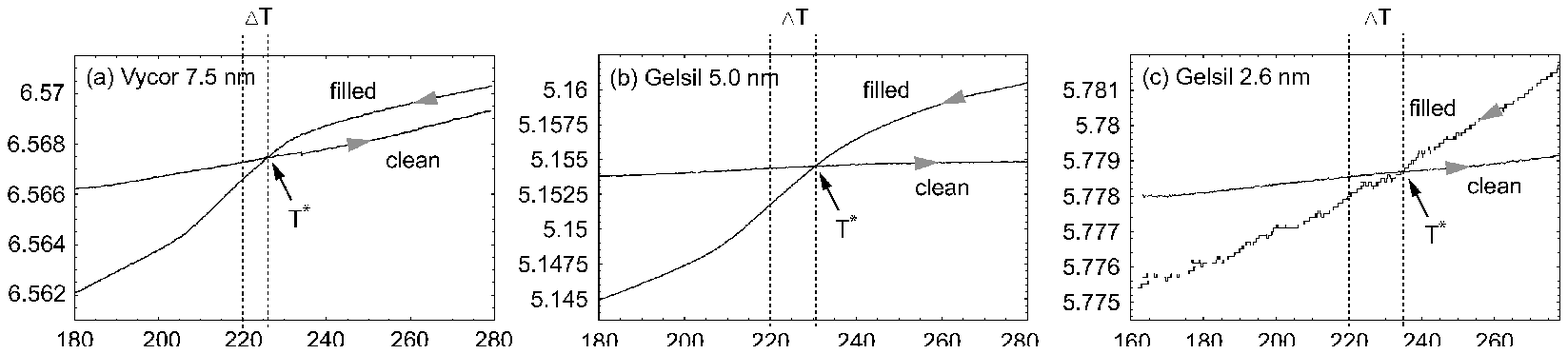}
\caption{Linear thermal expansion of empty and salol filled samples with pore diameters of (a) 7.5~nm, (b) 5.0~nm and (c) 2.6~nm.}
\label{fig:filled_vs_clean_overview}
\end{center}
\end{figure*}

\qquad  As Fig.~\ref{fig:filled_vs_clean_overview}a shows, for large pores and in a cooling process starting at RT, at higher temperatures the Vycor matrix is not affected by its filling. It contracts like the empty Vycor matrix with a thermal expansion coefficient $\alpha=\Delta h / (h\cdot\Delta T)=5.1\cdot10^{-6}\;\textrm{K}^{-1}$. Patkowski al.\cite{patkowski} proposed the possible flow and equilibration of the confined liquid well above $T_g$, which we also consider to be the case here. But as vitrification sets in at about 230 K, the filled Vycor matrix is subject to a contraction which is stronger compared to the empty Vycor sample. Strong interaction (H-bondings) between salol and the pore surface might be the reason for this. At smaller pores of filled Gelsil samples (Fig.~\ref{fig:filled_vs_clean_overview}b and c) additional contraction already starts at higher temperatures. For an estimation of the process developing negative pressure upon the filling liquid, the strain misfit between the glass and the host matrix is

\begin{equation}
\Delta \epsilon^{mf}=3 (\alpha_1-\alpha_{2})\Delta T
\end{equation}

with $\alpha_i$, the thermal expansion coefficients of the host matrix (1) and salol (2). Negative pressure then derives from $\Delta P=\Delta \epsilon / \kappa_T$, where $\kappa_T$ is the bulk compressibility of salol. The resulting shift of $T_g$, i.e.~

\begin{equation}
T_g(P)=T_g(P=0)\cdot\frac{\partial T_g}{\partial P}\Big|_{P=0}\cdot\Delta P
\end{equation}

crucially depends on the choice of $\Delta T$, the temperature range, in which the effective negative pressure upon salol develops. This effective temperature range can be estimated from our data as follows: As calculated from Eqn.~(\ref{eqn:filling_strain}) the host porous matrix expands with filling due to the negative capillary pressure which acts on the confined liquid. Since with cooling the liquid salol contracts, this stress relaxes and the composite is stress free if the filled sample height is the same as for the empty matrix which occurs at $T=T^{*}$ (see Fig.~\ref{fig:filled_vs_clean_overview}). So $\Delta T \approx T^{*}-T_g$. Results of these estimations are given in Tab.~\ref{tab:DeltaTg}. Parameters used for salol are $\kappa_T=5\cdot10^{-10} \;\textrm{Pa}^{-1}$ from Ref.~\onlinecite{compr_Salol}, the thermal expansion coefficient $\gamma_1=3\alpha_1=7.36\cdot10^{-4}\;\textrm{K}^{-1}$ from Ref.~\onlinecite{cukiermann}, and $\partial T_g/\partial P= 0.204 \;\textrm{K/MPa}$ from Ref.~\onlinecite{dTg_dP_Salol}. Our measurements are in accordance with enthalpy recovery results of S.~L.~Simon et al.\cite{enthalpy_recovery}. Their model shows that effective negative pressure develops 2 to 2.5~K above the reduced glass transition for samples with 11.6~nm and 25.5~nm pore sizes. Further, they state "...if negative pressure were the cause of the depressed $T_g$, the temperature at which isochoric conditions are imposed would have to be $\sim$ 20 to 40~K above $T_g$." For comparison we obtain a necessary $\Delta T=$ 10 to 40 K for $d=$ 7.5 to 2.6~nm pores, which is in very good agreement with Simon et al.

\begin{table}[h!]
\caption{Parameters of $\Delta T_g$ estimations, $\Delta T_g^{exp}=\Delta T_g^{np}+\Delta T_g^{conf}$.}  \label{tab:DeltaTg}
\begin{center}
\begin{tabular}{llll}
\qquad&\qquad Vycor&\qquad Gelsil5.0&\qquad Gelsil2.6\\
\hline \hline
$d$ (nm)&\qquad 7.5&\qquad 5.0&\qquad 2.6\\
$\alpha_2$ ($\textrm{K}^{-1}$)& \qquad $2.1\cdot10^{-5}$&\qquad $4.6\cdot10^{-5}$&\qquad $1.1\cdot10^{-5}$\\
$\Delta T$ ($\textrm{K}$)& \qquad 6 &\qquad 10 &\qquad 15 \\
$\Delta \epsilon^{mf}$ (\%)& \qquad -$0.40$ &\qquad -$0.60$&\qquad -$1.06$\\
$\Delta P$ (MPa)& \qquad -$8.1$ &\qquad -$12.0$& \qquad -$21.1$\\
\hline
$\Delta T_g^{np}$ (K)& \qquad -$1.6$ &\qquad -$2.4$&\qquad -$4.3$\\
$\Delta T_g^{conf}$ (K)& \qquad -$1.4$&\qquad -$2.9$&\qquad -$6.4$\\
\hline
$\Delta T_g^{exp}$ (K)& \qquad -$3.0$ &\qquad -$5.3$&\qquad -$10.7$\\
\hline
\hline
\end{tabular}
\end{center}
\end{table}

\begin{figure}
\begin{center}
\includegraphics{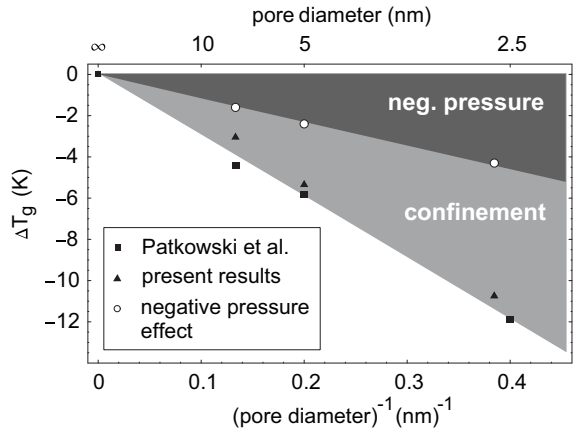}
\caption{Shift of glass transition temperature against $(\textrm{pore diameter})^{-1}$. Boxes are $T_gs$ from Fig.~\ref{fig:pore_center_tau}, triangles show literature values from Ref.~\onlinecite{patkowski}, open circles display the maximum negative pressure contribution (see chapter III.~C.~).}
\label{fig:DeltaTg_pore diameter}
\end{center}
\end{figure}

\qquad In our opinion our calculated $\Delta T_g^{np}$ is still overestimated for two reasons: First, using the bulk value $\alpha_2$ of the host matrix from Fig.~\ref{fig:filled_vs_clean_overview} does not take into account internal pore walls being affected by the negative pressure inside, relaxing to some extend and so reducing pressure. Secondly, thermal expansions of other glass forming liquids e.g.~toluene have been reported $1.5$ times smaller in confinement\cite{toulene1} compared to bulk.  Moreover, thermal expansion of liquid salol drops\cite{cukiermann} to a \textit{quarter} of its value at the glass transition. So, as the glass transition sets in, $\alpha_1$ starts to decrease and a purely pressure induced downshift $\Delta T_g$ would be even more diminished. Apart from this the reason for the size dependence of the thermal mismatch effect (see Fig.~\ref{fig:DeltaTg_pore diameter}, open circles) is not clear at all.

\section{Conclusions}

The glass transition of salol confined to porous host matrices of Vycor and Gelsil with pore sizes of 7.5, 5.0 and 2.6~nm has been measured for the first time by Dynamic Mechanical Analyzers (DMA 7 and Diamond DMA, Perkin Elmer). The dynamic complex elastic susceptibility data can well be fitted assuming two types of dynamic processes: A "bulk" relaxation in the core of the pores and a radially increasing "surface relaxation" of molecules near the pore surface. The calculated core relaxation time shows a typical Vogel-Fulcher temperature dependence and decreases with decreasing pore size $d$. This confinement induced acceleration of dynamics leads to a shift of the glass transition temperature $T_g\propto 1/d$, which is in perfect agreement with recent DSC results\cite{patkowski}. Measurements of the sample height with filling (adsorption swelling) and thermal expansion are used to calculate the effect of "negative pressure" due to thermal mismatch between the porous host matrix and the glass forming liquid. Such negative pressure could at least partly explain a shift of $T_g$ in confined glass forming liquids\cite{patkowski, review_Alcoutlabi, enthalpy_recovery}. Our data show that for salol this effect of thermal mismatch could describe at most 30\% of the observed downshift of $T_g$, which is in harmony with enthalpy recovery experiments\cite{enthalpy_recovery}.

\qquad In our opinion the main cause for the shift of $T_g$ is a hindering of cooperativity due to confinement. This is also supported by an estimation of this effect using the results of Hunt et al.\cite{hunt1}. They calculated the finite size effect of the glass transition from percolation and effective medium models, which yields

\begin{equation}
T_g(d)=T_g(\textrm{bulk}) - \frac{0.5\cdot E}{ln(t\cdot\nu_{ph})}\cdot\frac{r_0}{L}\;.
\label{eqn:hunt}
\end{equation}

Inserting $t=100$ s, $\nu_{ph}=1/\tau_0$, and our fit parameters from Tab.~\ref{tab:FitparameterVycorGelsil}, and assuming that the typical distance between molecules $r_0$ is about the diameter of a salol molecule\cite{salolsize2} ($d_0\approx0.8$~nm), we obtain $\Delta T_g^{Hunt}$ as 3.2, 4.8 and 9.1~K for 7.5, 5.0 and 2.6~nm pores, respectively. These calculated values agree surprisingly well with the measured confinement induced downshifts of $T_g(d)$  (see Fig.~\ref{fig:DeltaTg_pore diameter} and $\Delta T_g^{exp}$ in Tab.~\ref{tab:DeltaTg}).

\qquad Moreover Eqn.~(\ref{eqn:hunt}) predicts\cite{hunt1} that the size dependence of $\Delta T_g$ increases with increasing fragility\cite{Boehmer}
\begin{equation}
m=\frac{E\cdot T_g}{ln\;(10)\;(T_g-T_0)^2}\;,
\label{eqn:fragility}
\end{equation}

since $m\propto E$. Indeed, this correlation between $\Delta T_g(d)\sim m$ was verified experimentally for many systems, i.e.~for glycerol\cite{neg_pressure1} ($m=53$) $\Delta T_g$($d=2.5$ nm) $\approx -4$ K, benzyl-alcohol\cite{calorimetric2} ($m=65$) $\Delta T_g$($d=2.5$ nm) $\approx -9$ K, salol\cite{patkowski}($m=73$)$\Delta T_g$ ($d=2.5$ nm) $\approx -11$ K, o-terphenyl\cite{patkowski}($m=81$) $\Delta T_g$ ($d=2.5$ nm) $\approx -25$ K.

\qquad We think that these considerations, i.e.~the downshift of $T_g$ calculated via percolation theory, as well as the clear correlation between the magnitude of induced $T_g$ shift and the fragility of a glass forming liquid, both confirm our other findings (see Fig.~\ref{fig:DeltaTg_pore diameter}) that the main effect of the confinement is to suppress cooperative motion. Negative pressure effects although always present contribute only little.

\vspace{0.8cm}

{\bf Acknowledgements}: Support by the Austrian FWF (P19284-N20) and by the University of Vienna within the IC Experimental Materials Science ("Bulk Nanostructured Materials") is gratefully acknowledged. We thank Marie-Alexandra Neouze and the Institute of Materials Chemistry from the Vienna University of Technology for the $N_2$-characterization of our samples. We are grateful to J.~Bossy (CNRS Grenoble) for supplying us with Gelsil samples.

\end{document}